\documentstyle[12pt]{article}

\newcommand{\be}{\begin{equation}}
\newcommand{\ee}{\end{equation}}
\newcommand{\ba}{\begin{eqnarray}}
\newcommand{\ea}{\end{eqnarray}}

\newcommand{\eq}[1]{(\ref{#1})}
\newcommand{\Eq}{Eq.~\eq}

\newcommand{\al}{\alpha}
\newcommand{\tr}{\mathop{\rm tr}\nolimits}

\begin{document}

\title{\bf Quantum continuous measurements, \\
dynamical role of information \\
and restricted  path integrals}
\author{{\Large\bf Michael B. Mensky}\\[2mm]
P.N.Lebedev Physical Institute, 117924 Moscow, Russia}

\date{}

\maketitle

\begin{abstract}

The restricted-path-integral (RPI) theory of continuous quantum
measurements including the evolution of the measured systems and
phenomenon of decoherence is reviewed. The measured system is
considered as an open quantum system but without usage of any model of
the measurement (of the measuring medium or the system's environment).
The propagator of a measured system (conditioned by the measurement 
readout) is presented by RPI. In the important special case of 
monitoring an observable the propagator and the system's wave function 
satisfy Schr\"odinger equation with a complex Hamiltonian (depending 
on the measurement readout). Going over to the non-selective 
description of the measurement leads to the Lindblad master equation. 
In case of non-minimally disturbing measurements this gives theory of 
dissipative systems avoiding difficulties of other approaches. The 
whole theory is deduced from first principles of quantum 
mechanics. This proves that quantum mechanics includes theory of 
measurements and is therefore conceptually closed.

\end{abstract}

\section{Introduction}

Theory of quantum measurements is one of the most interesting and in a
sense mysterious topics in quantum mechanics. In last decades
essential progress in understanding quantum measurements was connected
with the concept of decoherence. The phenomenon of decoherence
consists in loss of quantum phases in the superposition existed before
the measurement, see \cite{Zeh-rev96} for a review. Decoherence
transforms an initial pure state (presented by a state vector, or wave
function) to a mixed state (presented by a density matrix). Continuous
quantum measurement (gradual decoherence) may be presented by a
time-dependent density matrix. In Markowian approximation this density
matrix satisfies a differential equation called master equation. 
Generic form of master equations was found by Lindblad 
\cite{Lindblad76}.

It is important that a quantum open system may be considered to be
continuously measured by its environment even if the latter is
not constructed for measurement. Therefore, theory of continuously
measured systems is in fact theory of open systems. We give here a
short review of theory of continuous quantum measurements based on
restricted path integrals (RPI).

Decoherence is caused by entanglement (quantum correlation) between
the measured system and its environment (measuring medium or
reservoir). The density matrix presents the result of decoherence in a
non-selective way when the measurement readout (final state of the
environment) is not known. The same physical process of measurement
may be presented in the selective way, by a state vector conditioned
by the measurement readout.

In case of a continuous measurement (gradual decoherence) the
time-dependent readout-conditioned state vector may be presented by
RPI which in Markowian approximation satisfies Schr\"odinger equation
with a non-Hermi\-te\-an Hamiltonian depending on the measurement
readout, see \cite{MBMbk00} for a review.

RPI approach to continuous measurement follows from first principles
of quantum mechanics in the form of Feynman's path integrals but may
also be derived by ordinary quantum-mechanical methods. The approach
is model-independent: RPI depends only on the information supplied by
the measurement but not on the concrete measuring medium. Some
features of RPI point out that the evolution of a continuously
measured (decohering) system is a fundamental type of evolution.
Particularly, RPI approach reveals the dynamical role of the
information escaping the system.

RPI differs from the usual Feynman path integral by a weight
functional in the integrand. Positive weight functionals describe
continuous measurements which result in minimal disturbance of the
measured system's state (given the information supplied by the
measurement). Those weight functionals which contain also phase 
shifts, determine non-minimally disturbing measurements leading to 
dissipation of the measured system.

\section{Continuous measurements and RPI}

Any measurement of a quantum system results from interacting this 
system with its environment. The direct way to describe the evolution 
of the measured system is to find the time-dependent density matrix of 
the closed system consisting of the system of interest and its 
environment and then trace out all degrees of freedom of the 
environment. The resulting reduced density matrix describes the 
measured system as an open system i.e. with the influence of the 
environment taken into account. This description is non-selective 
since all possible states of the environment (all possible measurement 
readouts) are accounted in it, with no selection of one of them. 
Examples of continuous measurements considered in this way may be 
found in \cite{Models}.

It is however possible to derive the evolution law of a continuously
measured system phenomenologically, making use of no explicit model of 
the environment and interaction with it. We shall show how this may be 
done with the help of restricted path integrals (RPI), see 
\cite{MBMbk00} for the review of the method. The resulting description 
will be selective, expressed in terms of a state vector (wave 
function) satisfying Schr\"odinger equation with the complex potential 
depending on the measurement readout. A non-selective description in 
the form of a master equation for the density matrix may be obtained 
then by integration over all possible measurement readouts.

\quad

According to Feynman's formulation of quantum mechanics, the
propagator (probability amplitude for transition from one point to
another) of a {\em closed} quantum system may be expressed in the form
of a path integral
\be
U_t(q'',q')
=\int d[p] \int_{q'}^{q''} d[q]\,
e^{\frac{i}{\hbar}\int_0^t (p\dot q - H(p,q,t)) dt}
\label{FeynPathInt}\ee
where the so-called {\em phase-space, or Hamiltonian, representation
of the path integral} is used. Integration must be performed in all
paths $[p]$ in the momentum space and in those paths $[q]$ in the
configuration space which have the given end points. The observables
$p$, $q$ may be multidimensional.

Details of the definition are not important for us here. The only
thing we need is that the propagator $U_t(q'',q')$ and the
corresponding time-dependent wave function $\psi_t(q'')$ satisfy
Schr\"odinger equation. If we introduce the evolution operator $U_t$
as an integral operator with the kernel $U_t(q'',q')$, then the
evolution of the system may be presented, correspondingly by the state
vector or density matrix, as follows:
\be
|\psi_t\rangle = U_t\, |\psi_0\rangle, \quad
\rho_t = U_t\, \rho_0\, U_t^{\dagger}.
\label{Evol}\ee

The ideology which was developed by Feynman as a background of
the path-integral approach is following. The probability amplitude for
the system to evolve along the given path is equal to the integrand in
\eq{FeynPathInt} (imaginary exponential of the action in units of the
Plank constant). However it is in principle impossible to know for a
closed system what path is chosen by it. Therefore, we have to sum up
the amplitudes corresponding to all paths, hence \Eq{FeynPathInt}
for the total probability amplitude (propagator).

If the system is continuously measured, this ideology directly leads
to RPI. Indeed, let a continuous measurement is performed on the
system during the time interval $[0,t]$. Then the measurement readout
supplies some information about the path chosen by the system in its
evolution. If $\al$ is the set of paths compatible with this
information, then the path integral has to be restricted onto the set
$\al$. The set $\al$ is called {\em quantum corridor}.

Instead, we may integrate over all paths but with the corresponding
weight functional in the integrand. This weight functional
$w_\al[p,q]$ has to be equal to unity for the paths $[p,q]$ belonging
to $\al$ and zero otherwise. Generalizing, we may describe the
measurement by any smooth functional $w_\al[p,q]$. Although no set of
paths can present this situation adequately, we may in this case speak
of the `quantum corridor with indistinct, fuzzy boundaries'. In fact, 
this is a more realistic description, since real measurements give 
information of just this type. $\al$ will be interpreted in this case 
either as a measurement readout or as a quantum corridor presented by 
$w_\al[p,q]$.

As a result, we have the following propagator for the system
undergoing a continuous measurement resulting in the measurement
readout $\al$:
\be
U_t^\al(q'',q')
=\int d[p] \int_{q'}^{q''} d[q]\, w_\alpha[p,q]\,
e^{\frac{i}{\hbar}\int_0^t (p\dot q - H(p,q,t)) dt} .
\label{PropMeasSys}\ee
The evolution of the system is then presented as follows:
\be
|\psi_t^{\alpha}\rangle = U_t^{\alpha} |\psi_0\rangle, \quad
\rho_t^{\alpha} = U_t^{\alpha} \rho_0
\left(U_t^{\alpha}\right)^{\dagger}
\label{EvolMeas}\ee

The system undergoing the measurement (therefore decohering) may thus
be presented by wave functions, or state vectors, but these wave
functions are conditioned by the measurement readouts $\al$.
Physically a measurement readout is nothing else than a state of the
environment, but here measurement readouts are presented by weight
functionals $w_\al[p,q]$ i.e. expressed in terms of paths of the
measured system. This is often advantageous.

Instead of a single propagator and a single unitary evolution operator 
for a closed system, a continuously measured system is presented by 
the set of {\em partial propagators} \eq{PropMeasSys} and {\em partial 
evolution operators} $U_t^{\alpha}$, one for each measurement readout 
$\al$. These operators are not unitary, so that the state vectors and 
density matrices \eq{EvolMeas} are not normalized. Instead, the square 
norm of the state vector or trace of the density matrix gives the 
probability density of the corresponding measurement readout:
\be
P(\al)=||\psi_t^\al||^2=\tr\rho_t^{\alpha}
\label{ProbMeasRead}\ee

The sum of the {\em partial density matrices} corresponding to all
possible measurement readouts gives the {\em total density matrix} 
\be
\rho_t = \int d\alpha\,  \rho_t^{\alpha}
=\int d\alpha\, U_t^{\alpha} \rho_0 \left( U_t^{\alpha}\right)^{\dagger}
\label{TotalDensMatr}\ee
presenting the same measurement non-selectively. The total density 
matrix $\rho_t$ should be normalized. This is the case if the {\em 
generalized unitarity condition} 
\be
\int d\alpha\,  \left( U_T^{\alpha}\right)^{\dagger}\, U_T^{\alpha} =
\bf 1
\label{GenUnitar}\ee
is fulfilled. \Eq{GenUnitar} is a condition on the weight functionals
$w_\al$ and the measure $d\al$ which provides consistency of the
definitions.

Of course, one may integrate over some subset of the measurement
readouts, providing a partially non-selective description of the
measurement. In practice only various partially non-selective
descriptions are realistic because the measurement readout $\al$ never
can be known precisely.

\section{Monitoring an observable}

Let us consider a special case of continuous measurements, 
monitoring an observable. An evident example is monitoring the
coordinate $q$. The measurement readout is then expressed by 
function $a(t')$, $t'\in [0,t]$, and interpreted as the statement that
at time $t'$ the coordinate $q$ has the value which differs from
$a(t')$ not more than by the entity $\Delta a$ characterizing the
measurement accuracy.
The same may be expressed as follows: the system evolves
along some path $[q]$ (in the coordinate space) which lies in the
corridor $\al$ of the width $\Delta a$ with the middle line $[a] =
\{a(t')|0\le t' \le t\}$. Such a measurement is described by RPI 
taken over the quantum corridor $\al$.

In generic case any observable $A(p,q)$ is monitored instead of the 
coordinate. The measurement readout expressed by the curve $[a] = 
\{a(t')|0\le t' \le t\}$ means that the value $A(t')=A(p(t'),q(t'))$ 
of the observable $A(p,q)$ at time $t'$ differs from $a(t')$ not more 
than by $\Delta a$ (the precision, or resolution, of the measurement). 
RPI must be taken over the set $\al$ of paths $[p,q]$ determined by 
these conditions. 

Of course, a more realistic description of the measurement is given by 
the corridor with fuzzy boundaries. It is expressed by a weight 
functional $w_\al[q]$ for monitoring the coordinate and by 
$w_\al[p,q]$ in the general case. The functional $w_\al[p,q]$ for 
monitoring $A(p,q)$ has to be approximately equal to unity if the 
curve $A(t')=A(p(t'),q(t'))$, $t'\in [0,t]$, is close to $a(t')$ and 
approximately equal to zero if these curves are far from each other.

The scale of closeness is $\Delta a$, but the definition of the 
`distance' between curves may depend on the concrete type of the 
measurement to be described. It is often reasonable to present 
monitoring by the Gaussian functional 
\be
w_{[a]}[p,q] = \exp\left(
{-\kappa \int_0^t [ A(t') - a(t') ]^2\,dt' }
\right)
\label{GaussWeightFunct}\ee
where $\kappa$ determines the strength, or resolution, of the 
measurement.

In principle $\kappa$ may depend on time, then it has to be in the 
integrand, but for permanent conditions of the monitoring $\kappa={\rm 
const}$. If the interval $t$ of measurement is fixed, the strength of 
the measurement $\kappa$ is connected with the `width of the corridor' 
$\Delta a$ by the equation $\kappa={1}/(t\,\Delta a^2)$, so that 
$\Delta a$ is the mean square deviation of the curve $A(t)$ from 
the middle line of the corridor $a(t)$. Therefore, for a constant 
strength of the measurement, $\Delta a$ decreases with time inversely 
proportional to $\sqrt{t}$.

The first, purely mathematical, reason to choose a Gaussian weight
functional is that this leads (in case of a quadratic Hamiltonian) to
Gaussian path integrals which may be precisely evaluated. There is
however a more deep physical reason. As is shown in the framework of a
special model \cite[Chapter~8]{MBMbk00} and is believed to be valid
generally, a Gaussian weight functional appears each time if the
continuous measurement consists of a large number of very weak short
measurements. This may be considered as a quantum version of the
Central Limiting Theorem from probability theory.

If the Gaussian weight functional \eq{GaussWeightFunct} is accepted
for monitoring, then RPI \eq{PropMeasSys} takes the form 
\begin{eqnarray}
U_t^{[a]}(q'',q')&=&\int d[p]\,d[q]\,
\exp\left\{ \frac{i}{\hbar} \int_0^t \big(p\dot q
- H(p,q,t)\big)\,dt \right.\nonumber\\
&-& \left.\kappa \int_0^T \big(A(p,q,t)-a(t)\big)^2 dt
\right\}
\label{RPIforMonitor}
\end{eqnarray}
This RPI is equal to Feynman path integral \eq{FeynPathInt} but with
the Hamiltonian
\be
H_{[a]}\,(p,q,t) = H(p,q,t) - i\kappa\hbar \,\big( A(p,q,t) - a(t)
\big)^2
\label{EffHamilt}\ee
containing (in comparizon with the initial Hamiltonian) an additional 
imaginary term. The propagator \eq{RPIforMonitor} and the 
corresponding conditioned wave function (state vector) 
$|\psi_t^{[a]}\rangle$ satisfy the effective Schr\"odinger equation 
\be
\frac{\partial}{\partial t} |\psi_t^{[a]}\rangle
  = \left[-\frac{i}{\hbar} \hat H
  -\kappa \,\Big(\hat A - a(t)\Big) ^2\right]\, |\psi_t^{[a]}\rangle
\label{EffSchrEq}\ee

The transition to the non-selective description of the measurement is
performed by summing up the partial density matrices over all
possible measurement readouts. In case of monitoring, this means
integrating $\rho_t^{[a]}$ over all curves $[a]$. The resulting
total density matrix may be shown to satisfy the following equation
(which is a special case of Lindblad equation):
\be
\frac{\partial}{\partial t} \rho_t = -\frac{i}{\hbar} [\hat H,\rho_t]
          - \frac 12 \kappa [\hat A, [\hat A, \rho_t]]
\label{NonSelectMasterEq}\ee

\section{Non-minimally disturbing monitoring}

We shall consider now a more general {\em non-minimally disturbing 
monitoring} \cite[Sect.~5.2.3]{MBMbk00}. It is described by a more 
general weight functional than \eq{GaussWeightFunct}. The functional 
\eq{GaussWeightFunct} damps out those paths which are not compatible 
with the information given by the monitoring. This leads to such a 
disturbance of the measured system's state which is unavoidable for 
the measurement supplying the given information. Besides this, some 
additional (non-minimal) disturbance may be performed during the 
measurement.

For example when the coordinate is measured, the momentum is
necessarily disturbed, but, besides this, the measurement of the 
coordinate may be accompanied by additional disturbance of the 
coordinate.

A non-minimally disturbing monitoring of observable $A(p,q)$ may be
described by a Gaussian weight functional similar to
\eq{GaussWeightFunct} but with an additional imaginary term in the
exponent. If we assume that this additional term is linear in the
measurement readout $a(t)$, then the weight functional is
\be
w_{[a]}[p,q]
=\exp \left\{ \int_0^t dt'
\left[ -\kappa \left( A-a(t')\right) ^2
-\frac i\hbar \left( \lambda \,a(t')B+C\right)
\right]
\right\}
\label{NonMinWeight}\ee
where $A=A(p,q)$, $B=B(p,q)$, $C=C(p,q)$ are arbitrary observables.
This type of measurement leads to dissipation of the measured system
\cite{MBMStenholm03}.

This may be shown as follows. If the partial evolution operator 
$U_t^{[a]}$ is RPI with weight functional \eq{NonMinWeight}, then the 
partial and total density matrices
\be
\rho_t^{[a]}=U_t^{[a]}\,\rho_0\, \left(U_t^{[a]}\right)^{\dagger},
\quad
\rho_t=\int d[a]\, \rho_t^{[a]}
\label{MonitorEvol}\ee
describe the measurement, correspondingly, selectively and
non-selectively.
It turns out \cite{MBMStenholm03} that the total density matrix
satisfies the master equation
\be
\frac {\partial \rho}{\partial t}
 =  -\frac i\hbar \left[ \hat{H}+\hat{C},\rho\right]
-\frac \kappa 2\left[ \hat{A},\left[ \hat{A},\rho\right] \right]
-\frac{\lambda ^2}{8\kappa \hbar ^2}
\left[ \hat{B},\left[ \hat{B},\rho\right] \right] 
 -\frac{i\lambda }{2\hbar }
\left[ \hat{B},\left[ \hat{A},\rho\right] _{+}\right] .
\label{a6}
\ee
(we omit index $t$ in $\rho_t$). With the notation 
$\hat{l}=\hat{A}-i\frac{\lambda}{2\kappa\hbar}\hat{B}$, we have 
\be
\frac {\partial \rho}{\partial t}
 =  -\frac i\hbar
\left[ \hat{H}+\hat{C}
-i\frac{\kappa\hbar}{4}\left( \hat{l}^{\dagger 2}-\hat{l}^2\right)\, ,
\, \rho\, \right]  
 - \frac{\kappa}{2} \left(
\hat{l}^{\dagger }\hat{l}\,\rho
-2\,\hat{l}\,\rho\,\hat{l}^{\dagger }
+\rho\,\hat{l}\,\hat{l}^{\dagger }\right)
\label{a8}
\ee
This equation is of Lindblad form \cite{Lindblad76} with a single
Lindblad operator $\hat l$. The original Hamiltonian is renormalized
by the measurement procedure.\footnote{The Lindblad master equation of
generic form, i.e. with a number of Lindblad operators, will result if
the corresponding number of observables undergo non-minimally
disturbing monitoring.}

From the physical point of view the measured system turns out to be
dissipative. For example, if $\hat H$ is a Hamiltonian of the harmonic
oscillator, $\hat A=\hat p$, $\hat B=\hat q$ and $\hat C=0$ (this
means that the oscillator's momentum is monitored and non-minimally
disturbed), then the resulting master equation
\be
\frac {\partial \rho}{\partial t}
 =  -\frac i\hbar \left[ \hat{H},\rho\right]
-\frac \kappa 2\left[ \hat{p},\left[ \hat{p},\rho\right] \right]
-\frac{\lambda ^2\omega ^2}{8\kappa \hbar ^2}
\left[ \hat{q},\left[ \hat{q},\rho\right] \right] 
 -\frac{i\lambda \omega }{2\hbar }
\left[ \hat{q},\left[ \hat{p},\rho\right] _{+}\right]
\label{a11}
\ee
is the well known master equation for a {\em Brownian motion} of the
harmonic oscillator \cite{Stenholm96}. The Brownian motion of the
oscillator is thus interpreted as the {\em effect of monitoring the
momentum} by a continuously acting environment (reservoir). The
friction coefficient of the oscillator is
$\gamma={2\hbar\kappa}/{m\omega}$.

\section{Conclusion}

We reviewed here RPI approach to theory of continuous quantum
measurements \cite{MBMbk00} which is especially important since it may 
be considered as theory of decoherence of open quantum systems in 
general \cite{PIF03}.

Being model-independent, this approach is efficient in various
applications (e.g. measurements on harmonic oscillators, visualization 
of a level transition and measurements of quantum fields). Besides, 
this approach makes evident fundamental features of the phenomenon of 
decoherence.

The most important of these features is the dynamical role of 
information: the evolution of a continuously measured (decohering) 
system does not depend of details of the measuring medium, but is 
determined by the measurement readout, i.e. by the information 
recorded in the environment.

One more remarkable feature is that RPI theory of measurement follows 
from first principles of quantum mechanics (from Feynman path-integral 
theory). Therefore, quantum mechanics includes theory of measurements 
and, contrary to the wide-spread opinion, is conceptually closed. The 
only unsolved conceptual problems in quantum mechanics are those 
connected with the role of consciousness in quantum measurements 
\cite{mbmUFN2000}. 

Presentation of the measured (decohering) open systems by RPI is 
universal. Although monitoring may be presented simpler, by 
Schr\"odinger equation with a complex Hamiltonian, but this is 
impossible for the measurements which are `integral in time'. Even 
monitoring turns out to be integral in time in the non-Markowian 
approximation: the number $a(t')$ is actually not the value of $A$ at 
time $t'$ but (owing to inertial properties of the measuring medium) 
the average of the values of $A$ over some period containing 
$t'$. The weight functional is then more complicated than 
\eq{GaussWeightFunct} or \eq{NonMinWeight} \cite[Sect.~5.3]{MBMbk00} 
and RPI cannot be reduced to a differential equation.

\end{document}